\documentclass[letterpaper,11pt,fleqn]{article}
\pdfoutput=1
\usepackage{jheppub}

\usepackage{graphicx}
\usepackage[figuresright]{rotating}
\usepackage{bm,amsmath,amssymb}
\usepackage[mathscr]{eucal}
\usepackage{color}

\long\def\comment#1{ }
\newcommand{\eqn}[1]{Eq.~\eqref{#1}}
\newcommand{\beq}{\begin{equation}}
\newcommand{\eeq}{\end{equation}}
\newcommand{\nn}{\nonumber\\}
\newcommand{\dif}{{\rm d}}
\newcommand{\rmd}{{\rm d}}
\newcommand{\rme}{{\rm e}}
\newcommand{\rmi}{{\rm i}}
\newcommand{\rmP}{{\rm P}}

\newcommand{\rmTr}{{\rm Tr}}

\newcommand{\del}{\partial}

\newcommand{\order}[1]{\mcal{O}{\left(#1\right)}}
\newcommand{\mcal}{\mathcal}
\newcommand{\wt}{\widetilde}

\newcommand{\bq}{\bm{q}}
\newcommand{\bp}{\bm{p}}
\newcommand{\bx}{\bm{x}}
\newcommand{\by}{\bm{y}}

\newcommand{\bz}{\bm{z}}

\newcommand{\br}{\bm{r}}

\newcommand{\abar}{\bar{\alpha}_s}

\newcommand{\sdla}{{\rm \scriptscriptstyle DLA}}

\newcommand{\minus}{\!-\!}
\newcommand{\Nc}{N_{\rm c}}
\newcommand{\CF}{C_{\rm F}}
\newcommand{\Nf}{N_{\rm f}}

\title{\Large Collinearly improved JIMWLK evolution in Langevin form} 

\author[a]{Yoshitaka Hatta}

\author[b]{and Edmond Iancu}

\affiliation[a]{Yukawa Institute for Theoretical Physics, Kyoto University, Kyoto 606-8502, Japan}

\affiliation[b]{Institut de physique th\'{e}orique, Universit\'{e} Paris Saclay, CNRS, CEA, F-91191 Gif-sur-Yvette, France}

\emailAdd{hatta@yukawa.kyoto-u.ac.jp}
\emailAdd{edmond.iancu@cea.fr}

\abstract{The high-energy evolution of Wilson line operators, which at leading order is
described by the Balitsky-JIMWLK equations, receives large radiative corrections 
enhanced by single and double collinear logarithms at next-to-leading order and beyond.
We propose a method for resumming such logarithmic corrections to all orders, at the level
of the Langevin formulation of the JIMWLK equation. The ensuing, collinearly-improved
Langevin equation features generalized Wilson line operators, which depend not only upon 
rapidity (the logarithm of the longitudinal momentum), but also upon
the transverse size of the color neutral projectile to which the Wilson lines belong.
This additional scale dependence is built up during the evolution, 
via the condition that the successive emissions of soft gluons be ordered in time. 
The presence of this transverse scale in the Langevin equation furthermore allows for the
resummation of the one-loop running coupling corrections.
}


\begin{document}
\maketitle
\section{Introduction}
\label{sec:Intro}
It is by now well established that the Wilson lines are the proper degrees of freedom for
describing scattering in QCD at sufficiently high energies. For instance, in the collision
between a dilute projectile and a dense target --- such as in deep inelastic scattering
at small Bjorken $x$, or in proton-nucleus collisions at the LHC ---, the forward scattering amplitude 
can be computed by associating a Wilson line to each of the partons composing the projectile. 
Similarly, to compute the (partonic) cross-section for particle production in a dilute-dense collision, 
one must associate Wilson lines to all the partons that are produced in the final state ---
separately in the amplitude and in the complex-conjugate amplitude. A Wilson line is a path-ordered 
phase built with the color field of the target which describes the change in the wavefunction
of a parton (notably, its color precession) due to its multiple scattering off the target, 
as computed within the eikonal approximation. To obtain the physical amplitudes or cross-sections, 
one must also average the (gauge-invariant) product of Wilson lines describing the scattering
`in a given event' (i.e. for a given configuration of the color fields in the target)
over all such configurations. The effective theory for the 
{\em Color Glass Condensate} (CGC) provides an appropriate weight function --- a functional 
probability distribution --- to that aim \cite{Iancu:2002xk,Iancu:2003xm,Gelis:2010nm}.

This overall picture is subjected to quantum evolution. The most important quantum
corrections in the high-energy regime of interest are those associated with the emission
of {\em soft gluons}, i.e. gluons which carry only a small fraction $x\ll 1$ of the longitudinal
momentum of their parent parton. In the leading logarithmic approximation (LLA), in which one
keeps only those perturbative corrections where each power of $\alpha_s$ is accompanied
by a `rapidity' logarithm $Y\equiv\ln(1/x)$, the high-energy evolution of the Wilson lines
is described by the Balitsky--JIMWLK equations\footnote{The acronym `JIMWLK' stands for Jalilian-Marian, Iancu, McLerran, Weigert, Leonidov and Kovner.}
\cite{Balitsky:1995ub,JalilianMarian:1997jx,JalilianMarian:1997gr,Kovner:2000pt,Iancu:2000hn,Iancu:2001ad,Ferreiro:2001qy}. Albeit equivalent, the Balitsky hierarchy and the JIMWLK equation
have been obtained by studying different physical processes and via different manipulations.
The Balitsky equations describe the coupled evolution of scattering amplitudes for dilute projectiles.
It was derived \cite{Balitsky:1995ub} by studying the evolution of the projectile wavefunctions in the
background of the strong color field of the Lorentz-contracted target (the `shockwave').
Each step of this evolution adds a softer gluon (that can be emitted by any of the preexisting
partons) and thus introduces a new Wilson line in the adjoint representation which describes
the eikonal scattering of that softer gluon. The JIMWLK equation, on the other hand, governs
the evolution of the multi-gluon correlations in the wavefunction of the dense target, in the presence
of the non-linear effects associated with large gluon occupation numbers. This is a {\em functional}
equation, in that it refers to the evolution of the CGC weight function --- the functional 
probability distribution alluded to above.
In terms of Wilson lines, the Balitsky hierarchy depicts the multiplication of Wilson lines representing
the projectile, while the JIMWLK equation describes the change in the target field which enters 
the Wilson lines, for a fixed structure of the projectile.

This being said, the JIMWLK Hamiltonian can be also used to act directly on (gauge-invariant)
products of Wilson lines, in which case it reproduces the Balitsky hierarchy. But the scope of the
JIMWLK evolution is somewhat broader, as this is not process-dependent ---
it refers to the gluon distribution in a dense target independently of any scattering process.
As such, it can also be used to construct the `initial conditions' for ultrarelativistic nucleus-nucleus
collisions, that is, the color charge distributions in the incoming nuclei. 
Besides, the fact that the JIMWLK Hamiltonian has a Fokker-Planck structure (a second-order,
functional, differential operator with a factorized kernel) allows for an equivalent reformulation
as a functional Langevin process  \cite{Weigert:2000gi,Blaizot:2002xy}, 
which is crucial for the feasibility of numerical studies. So far, the only
exact solutions to the Balitsky hierarchy were obtained via numerical simulations of this
Langevin process \cite{Rummukainen:2003ns,Lappi:2011ju,Dumitru:2011vk,Lappi:2012vw,Schlichting:2014ipa,
Schenke:2016ksl}.
(All the other studies of this evolution have involved additional approximations,
like the multi-color limit $N_c\gg 1$, in which the Balitsky-JIMWLK hierarchy reduces to the 
Balitsky-Kovchegov (BK) equation \cite{Balitsky:1995ub,Kovchegov:1999yj}, 
or various mean approximations to the CGC weight function 
\cite{Iancu:2002aq,Blaizot:2004wv,Kovchegov:2008mk,Marquet:2010cf,Dominguez:2011wm,Iancu:2011ns,Iancu:2011nj,Lappi:2016gqe}.)
Importantly, this Fokker-Planck structure supports the probabilistic 
interpretation of the CGC effective theory: it ensures that the CGC weight
function remains positive definite and properly normalized under the evolution.

Whereas conceptually limpid and technically under control, the LLA is however
insufficient for applications to physics: the evolution that it predicts is way too fast 
(say, in terms of the growth of the gluon distribution prior to saturation, or of the saturation
momentum, with increasing energy) to agree with the phenomenology. This has motivated
strenuous efforts towards computing the next-to-leading order (NLO) corrections 
\cite{Triantafyllopoulos:2002nz,Kovchegov:2006wf,Kovchegov:2006vj,Balitsky:2006wa,Balitsky:2008zza,Avsar:2011ds,Balitsky:2013fea,Kovner:2013ona}, which eventually 
established the NLLA versions of the BK equation  \cite{Balitsky:2008zza} and of
the Balitsky-JIMWLK hierarchy \cite{Balitsky:2013fea,Kovner:2013ona}.
However, when taken at face value, these NLLA equations are rather disappointing:
the NLO corrections include terms which are negative and potentially large, since enhanced
by double and single `collinear logarithms', i.e. logarithms of the ratio between the characteristic
transverse momenta of the projectile and the target. When this ratio is large, as is typically the case
in dilute-dense collisions, these corrections render the NLLA approximation unstable \cite{Lappi:2015fma}
and hence useless in practice. A similar problem had been originally identified in relation to
the NLO version \cite{Fadin:1995xg,Fadin:1997zv,Camici:1996st,Camici:1997ij,Fadin:1998py,Ciafaloni:1998gs}
of the BFKL equation \cite{Lipatov:1976zz,Kuraev:1977fs,Balitsky:1978ic}  (the linearized version of the BK equation valid when the scattering is weak) and several methods have been 
devised to deal with it in that context
 \cite{Kwiecinski:1997ee,Salam:1998tj,Ciafaloni:1999yw,Altarelli:1999vw,Ciafaloni:2003rd,Vera:2005jt}.
Namely, it has been understood that collinear logs should generically occur in higher
orders, due to the interplay between the BFKL and the DGLAP evolutions, and that their
resummation is compulsory in order to restore the convergence and the predictability
of the perturbative expansion. But the original proposals in that sense  
\cite{Kwiecinski:1997ee,Salam:1998tj,Ciafaloni:1999yw,Altarelli:1999vw,Ciafaloni:2003rd,Vera:2005jt}
were specially tailored for the linear, BFKL, equation (in particular, they were formulated in Mellin space)
and cannot be directly transposed to the non-linear, Balitsky-JIMWLK, evolution.

More recently, new resummation methods have been proposed 
\cite{Beuf:2014uia,Iancu:2015vea,Iancu:2015joa} which are directly formulated
in transverse coordinate space and thus can naturally accommodate non-linear effects
like multiple scattering in the eikonal approximation. These methods focus on the
high-energy evolution of the dilute projectile.  So far they have been restricted to the BK equation,
that is, to a projectile which is a color dipole and to the large-$N_c$ limit. In Ref.~\cite{Beuf:2014uia}
it has been argued that the double collinear logarithms can be faithfully resummed by enforcing
time-ordering in the sequence of soft gluon emissions (see also
\cite{Ciafaloni:1987ur,Andersson:1995ju,Kwiecinski:1996td,Salam:1998tj,Motyka:2009gi,Avsar:2009pf} for  earlier, similar, ideas). This prescription leads to an improved version of the BK equation
which is non-local in rapidity ($Y$). An explicit diagrammatic calculation justifying this argument
has been presented in Ref.~\cite{Iancu:2015vea}. Moreover, it was shown
there that the non-local equation with time-ordering 
can be equivalently rewritten as a {\em local} equation, but with 
modified kernel and initial conditions which resum double collinear logarithms to all orders.
Subsequently, this resummation has been extended to include a subset of the single collinear logs
and  the running coupling corrections \cite{Iancu:2015joa}. Numerical
solutions to the ensuing, {\em collinearly-improved}, version of the BK equation
\cite{Iancu:2015vea,Iancu:2015joa,Albacete:2015xza,Lappi:2016fmu} demonstrate the
effects of the resummations in stabilizing and slowing down the evolution.
The evolution speed (or `saturation exponent') extracted from these solutions
is now in good agreement with the phenomenology. And indeed, based on these improved equations, 
one was able to obtain very good fits to the HERA data for deep inelastic scattering at
small Bjorken $x\le 0.01$ with only few free parameters (which refer to 
the initial conditions at low energy) \cite{Iancu:2015joa,Albacete:2015xza}.

It is {\em a priori} unclear if, and how, the resummation schemes in 
\cite{Beuf:2014uia,Iancu:2015vea,Iancu:2015joa} can be 
extended to the full Balitsky-JIMWLK hierarchy, i.e. beyond the large-$N_c$ approximation.
Besides allowing for a study of the finite-$N_c$ corrections, such an extension would 
give access to more complicated scattering operators,
such as the color quadrupole, which occur in the calculation of multi-particle 
correlations in dilute-dense collisions (see e.g. \cite{Dominguez:2011wm}),
and it would provide the initial color charge distributions for ultrarelativistic 
nucleus-nucleus collisions. It could also bring more insight into the structure of
perturbative QCD at high energies. Last but not least, this could provide inspiration
for collinear-improvement in a different physics problem, namely in the context of 
the BMS equation \cite{Banfi:2002hw}
and its finite--$N_c$ extension \cite{Weigert:2003mm,Hatta:2013iba},  which govern the resummation of
non-global logarithms in the evolution of jets with kinematical vetoes (see \cite{Neill:2015nya}
for a recent discussion and more references). Notice also that, in order to be truly useful, such an
extension should also admit a stochastic formulation, say, as a functional Langevin 
equation for the Wilson lines, and thus lend itself well to numerical simulations. 

There are two different ways to appreciate the potential difficulties with such
an extension (see also Sect.~\ref{sec:conc} below for a more detailed discussion). 
If one attempts to build a collinearly-improved version of the JIMWLK
Hamiltonian, then one should take guidance from the {\em local} version
of the BK equation constructed in Refs.~\cite{Iancu:2015vea,Iancu:2015joa}.
(We recall that the rapidity $Y$ plays the role of an evolution time, hence an
equation non-local in $Y$, like that presented in \cite{Beuf:2014uia}, cannot be 
derived via a Hamiltonian principle.) However, by inspection of the results in 
\cite{Iancu:2015vea,Iancu:2015joa}, it is quite clear that the resummed BK kernel
does not have the proper structure to be obtainable from a Hamiltonian
of the Fokker-Plank type (namely, it cannot be factorized as the product of
two independent emissions; see Sect.~\ref{sec:conc} for details). This strongly
suggests that it should be impossible to construct a {\em local} (in $Y$)
Langevin equation with collinear improvement. It furthermore rises doubts
about the possibility to maintain a probabilistic interpretation for the CGC
effective theory beyond leading order. 

If, on the other hand, one attempts to follow the strategy  in \cite{Beuf:2014uia}
--- that is, to construct a {\em non-local} Langevin equation which is endowed with
time ordering ---, then one immediately faces the following 
problem: the kinematical constraints expressing time-ordering involve the transverse
sizes (or momenta) of the parent projectile and of
the daughter gluons; yet, this transverse-scale dependence is not encoded 
in the usual definition of the Wilson lines operators.

In this paper, we shall identify a strategy to circumvent this last problem; that is,
we shall propose a generalization of the Wilson line operator
which `knows' about the transverse size of the parent projectile.
In turn, this will allow us to construct a non-local version of the
Langevin equation where time-ordering is enforced via kinematical constraints.
As we shall see, the additional scale dependence of the Wilson lines
is naturally and automatically built up during the evolution,
via the kinematical constraints. As a result, our Wilson lines will depend upon 3 independent
variables: the transverse coordinate $\bx=(x^1,x^2)$ of the corresponding parton, 
the rapidity $Y$, and the transverse size $r$ of the bare projectile which
initiates the evolution. The dependence upon $\bx$ already exists at tree-level, while
the two other ones are introduced by the high-energy evolution with time-ordering.
For a more complicated projectile, like a quadrupole, which involves several transverse 
scales --- the transverse separations between the `valence'  partons ---,
our construction applies so long as these various scales are commensurable with each other. 
Their precise values are unimportant since the scale dependence which is ultimately built 
in the observable (the projectile $S$-matrix) is merely logarithmic.  Indeed, the
effect of the time-ordering within the Langevin equation is to resum to all orders
the radiative corrections enhanced by double collinear logarithms. 

This additional scale dependence of the Wilson line operators will also
enable us to resum a particular class of single collinear logarithms
together with the running coupling corrections. Indeed, the transverse size 
of the projectile sets the scale for the collinear logs and for the
running of the coupling\footnote{Previous proposals for introducing a running coupling 
within the leading-order JIMWLK equation \cite{Lappi:2011ju,Dumitru:2011vk,Lappi:2012vw}
were not fully satisfactory precisely because of this problem:
the lack of information about the projectile transverse size.} 
(in the case where the daughter gluons have lower transverse
momenta). Our strategy for these additional resummations is similar to that followed 
in \cite{Iancu:2015joa} in relation with the BK equation. 

Our main new results are presented in Eqs.~\eqref{Langevincoll}--\eqref{rightcoll}
and in  Eqs.~\eqref{leftfin}--\eqref{rightfin}. The first set of equations expresses the Langevin
evolution with time-ordering and running coupling; the second set of equations encompasses the
additional resummation of single collinear logarithms. 
It is this last set of equations that should be numerically solved for applications to the
phenomenology. Clearly, the additional transverse-scale dependence complicates the 
numerical problem, by enlarging the associated functional space. We nevertheless believe that
this problem is tractable.

The paper is organized as follows. In Sect.~\ref{sec:LO}, we succinctly review the 
Langevin formulation of the leading-order JIMWLK evolution, with emphasis on its 
physical interpretations as either target evolution, or as the evolution
of the projectile. A good understanding of the physical picture is indeed important, 
as it will guide us when introducing time-ordering, later on. Then, in Sect.~\ref{sec:TO}, we recall
the origin of the radiative  corrections enhanced by double-collinear logarithms and, notably,
their relation with the condition of time-ordering for successive gluon emissions in the
`hard-to-soft' evolution of the projectile. We also clarify a rather subtle point concerning the emergence
of such corrections in the context of the JIMWLK equation. (This was originally
devised as `soft-to-hard' target evolution and one may think that it should be free of double collinear 
logs; this is however not the case for reasons to be explained at the end of Sect.~\ref{sec:TO}.)
Sect.~\ref{sec:COLL} is the main section, which contains our new results. Besides the two versions
of the collinearly-improved Langevin equation already mentioned, cf. 
Eqs.~\eqref{Langevincoll}--\eqref{rightcoll} and respectively Eqs.~\eqref{leftfin}--\eqref{rightfin},
we also present the version of the BK equation emerging from our
Eqs.~\eqref{Langevincoll}--\eqref{rightcoll} and compare this version with the one proposed
in \cite{Beuf:2014uia}. Finally, in Sect.~\ref{sec:conc} we summarize our results and extract
from them some lessons for the JIMWLK evolution beyond leading order.

\section{The Langevin formulation of the leading-order JIMWLK evolution}
\label{sec:LO}

Our starting point is the Langevin formulation \cite{Weigert:2000gi,Blaizot:2002xy}
 of the leading order (LO) JIMWLK equation
\cite{JalilianMarian:1997jx,JalilianMarian:1997gr,Kovner:2000pt,Iancu:2000hn,Iancu:2001ad,Ferreiro:2001qy}. This involves 
a functional stochastic equation on the SU$(N_c)$ group manifold which describes the 
high-energy evolution of the Wilson lines in the leading logarithmic approximation (LLA).
The `Wilson lines' are unitary matrices expressing the color precession of an elementary
projectile (a quark or a gluon) which scatters off a (generally strong) color background field
representing the gluon distribution of a dense target. It is understood that the partonic projectile
belongs to a color-neutral system, such as a quark-antiquark dipole, which is built with 
several such partons and which remains dilute on all the resolution scales explored by the collision
and by its high-energy evolution. The ultimate goal is to compute the elastic $S$--matrix for the scattering
between that dilute projectile and the dense target, in the eikonal approximation.

 The high-energy evolution proceeds via
the additional emission of soft gluons (i.e. gluons which carry a small fraction of the longitudinal 
momentum of their parent parton) when increasing the rapidity separation $Y=\ln(s/Q_0^2)$
between the projectile and the target. Here, $s$ is the center-of-mass energy squared 
and $Q_0$ is the characteristic transverse scale of the target, e.g. its saturation momentum at
the rapidity $Y_0=0$ at which one starts the evolution. The LLA is valid when $\alpha_s Y\gtrsim 1$ 
and consists in resumming the radiative corrections of the type $(\alpha_s Y)^n$ to all orders. 

As generally with stochastic equations, the Langevin formulation of the JIMWLK evolution
requires a discretization of the `evolution time', here the rapidity $Y$, so we shall
write $Y=n\epsilon$,
with integer $n\ge 0$. Also, we shall use the `symmetric' version of the Langevin equation
\cite{Iancu:2011nj,Lappi:2012vw}, where
the change in a Wilson line associated with one evolution step ($n\to n+1$) involves both 
`left' and `right', infinitesimal, color precessions, whose physical meaning will be shortly explained.
Then the Langevin equation reads (for a Wilson line in the fundamental representation,
for definiteness): 
\beq\label{LangevinLO}
U_{(n+1)\epsilon}^\dagger(\bx) = \exp \left[\rmi\epsilon \alpha^L_{n+1}(\bx)\right]
 U^\dagger_{n\epsilon}(\bx)\exp\left [-\rmi\epsilon \alpha^R_{n+1}(\bx)\right ]\,,
\eeq
where $\bx$ is the transverse coordinate of the quark projectile, which is not modified
by the scattering in the eikonal approximation. The other notations are as follows:
\begin{eqnarray}
\alpha^L_{n+1}(\bx)& \equiv&\frac{\sqrt{\alpha_s}}{\pi}\int_{\bz} {\mathcal K}_{\bx \bz}^i \nu_{n+1,\bz}^{ia}t^a\,, \label{left} \\
\alpha^R_{n+1}(\bx)&\equiv&\frac{ \sqrt{\alpha_s}}{\pi}\int_{\bz}  {\mathcal K}_{\bx \bz}^i \nu_{n+1,\bz}^{ia} \wt{U}^{\dagger ab}_{n\epsilon}(\bz)t^b\,, \label{right}
\end{eqnarray}
where the $t^a$'s, with $a=1,\dots,N_c^2-1$, are the generators of the SU$(N_c)$ Lie algebra in the fundamental representation, $\wt{U}_{n\epsilon}$ is a Wilson line in the adjoint representation which
describes the color precession of the emitted gluon, and 
\beq\label{Kdef}
{\mathcal K}_{\bx \bz}^i\equiv\frac{(\bx-\bz)^i}{(\bx-\bz)^2}
\,=\,-{\rmi}\int\frac{\rmd^2\bp}{2\pi}\,\frac{p^i}
 {p_\perp^2}\ \rme^{\rmi \bp\cdot(\bx-\bz)}
\eeq
is the Weizs\"{a}cker-Williams kernel for the emission of a soft gluon (the propagator
of the emitted gluon in the transverse plane; see also below). Finally, $ \nu_{n,\bz}^{ia}$
is a Gaussian white noise, with correlator
\beq
\langle \nu_{m,\bx}^{ia}\nu_{n,\by}^{jb}\rangle = \frac{1}{\epsilon}\,\delta^{ij}\delta^{ab}\delta_{mn}\delta_{\bx\by}\,. \label{noise}
\eeq

To better understand this evolution, let us first recall the generic structure of the Wilson
line describing the eikonal scattering between a quark projectile and a dense target:
\begin{align}\label{Udef}
 U^{\dagger}(\bx) = \rmP \exp\left[\rmi g \int \dif x^+ A^-_a(x^+,\bx) t^a\right].
 \end{align}
We work in a Lorentz frame where the dilute projectile is a right mover, while the target is a left mover.
Hence, the light-cone (LC) coordinate $x^+$ plays the role of a `time' for the projectile and, respectively, 
a longitudinal coordinate for the target. The quark propagates along the positive LC, $x^3=x^0$, or
$x^-=0$; its trajectory is parametrized by $x^+$ and the fixed transverse coordinate $\bx$. The target
is a shockwave localized near $x^+=0$. The interaction consists in the eikonal coupling between the 
quark color current and the LC component $A^-_a$ of the target color field. It results in the Wilson line
\eqref{Udef}, which physically describes the rotation of the quark color state. (The symbol P in 
\eqref{Udef} stands for time ordering: with increasing $x^+$, matrices are ordered from right to left.)
The `background' field $A^-$ is frozen during a given collision, by Lorentz time dilation, but it can 
randomly vary from event to event and physical quantities are obtained after averaging over $A^-$. 
This averaging must be performed at the level of the gauge-invariant product of 
Wilson lines which represents a color-neutral projectile, such as a dipole. Within the CGC 
effective theory \cite{Iancu:2002xk,Iancu:2003xm,Gelis:2010nm},
this average is computed using the `CGC weight function', a functional probability
distribution which evolves with $Y$ according to the JIMWLK equation. Alternatively, and equivalently,
the LO JIMWLK evolution of the CGC weight function can be reformulated as the stochastic
equation \eqref{LangevinLO} for Wilson lines with open color indices.
As it should be clear from the above, we work in a gauge where the component 
$A^-_a$ is non-zero. For the physical picture below, the most useful such gauge is the
projectile LC gauge $A^+_a=0$.

The physical interpretation of \eqn{LangevinLO} can be given in terms of 
either target or projectile evolution. For more clarity, we shall briefly describe 
here both points of views. But the
perspective of projectile evolution turns out to be more useful for our new developments
in this paper.

When the rapidity increment $\rmd Y=\epsilon$ is used to boost the target,
\eqn{LangevinLO} describes the evolution of the color field $A^-$ due to the
emission of a soft gluon by color sources in the target. This new gluon 
carries a smaller fraction of the target longitudinal momentum $P^-$, hence it is typically
delocalized over larger values of $|x^+|$. This is consistent with \eqn{LangevinLO}, 
which shows that the field $A^-$ evolves by extending its support towards larger
values of $|x^+|$  : the `left' and `right' color rotations in \eqn{LangevinLO}
add new layers to $A^-$, which are located at larger (and positive) and, respectively,
smaller (and negative) values of $x^+$ as compared to the color field from the previous step. 
It furthermore evolves by developing new correlations,
as introduced by the noise term. Physically, the noise term represents the
color charge density of the emitted gluon. The adjoint Wilson line $\wt{U}_{n\epsilon}$ 
in \eqn{right} describes the color precession of the noise (i.e. of the emitted gluon)
by the color field built in the previous steps. Via this mean field effect, the correlations
built by iterating \eqn{LangevinLO} are non-linear to all orders in the color field (or gluon)
distribution in the target.

When \eqn{LangevinLO} is interpreted as the evolution of the projectile, the color
field of the target is not evolving anymore --- for any rapidity separation $Y$,
this is still distributed according to the CGC weight function at the original
rapidity $Y_0=0$ --- but the $S$--matrix of the
projectile changes due to soft gluon emissions within the projectile wavefunction.
Accordingly, the unitary matrix $U_{n\epsilon}^\dagger(\bx)$ should not be viewed
anymore as a {\em genuine} Wilson line --- if by a `genuine Wilson line' we understand
the path-ordered phase built with the field $A^-$ according to \eqn{Udef} ---, but
rather as a more complicated scattering operator, which refers to 
a multi-partonic system built with the original quark at $\bx$
together with the $n$ soft gluons generated by  
the evolution. These gluons are strongly ordered in $k^+$ 
--- the longitudinal momentum for a right-mover --- which decreases
from the projectile towards the target.
This being said, we shall still refer to the unitary operator 
$U_{n\epsilon}^\dagger(\bx)$ as a `Wilson line',  for brevity, while keeping in mind that, 
when $n\ge 1$, this is not a {\em bare} Wilson line anymore.

Consider now an additional step in the evolution, $n\to n+1$, in which an even softer gluon
is being emitted --- either {\em before} the collision 
with the shockwave (the infinitesimal color precession on the right of $U_{n}^\dagger(\bx)$), 
or {\em after} the collision (the color precession on the left). If the emission occurs prior
to the collision, then the soft gluon itself will eventually cross the shockwave and 
thus acquire a color precession described by $\wt{U}_{n}(\bz)$. 
One may expect this additional gluon to be radiated by any of the
$n+1$ color sources generated in the previous steps, 
but this is not the content of \eqn{LangevinLO}.
Rather, this equation describes a process where the soft gluon is emitted by the quark
at $\bx$ alone. This is clear e.g. from the fact that 
Eqs.~\eqref{left}--\eqref{right} involve the Weizs\"{a}cker-Williams propagators from $\bx$
(the emission point) to the point $\bz$ where the gluon is measured by the noise 
term\footnote{Within the projectile evolution, the noise plays the role of a `detector' which measures the soft gluon in the final state, after the scattering has been completed.}. This reflects the fact that 
\eqn{LangevinLO}  describes the {\em backward} evolution of the projectile.

At this point, it is useful to 
briefly remind the difference between the `forward' and the `backward' evolutions.
In the forward evolution, the rapidity increment is viewed as a decrease in the rapidity of the 
softest gluons from the projectile that can be resolved by the target. This opens the phase-space
for the emission of even softer gluons, from any of the preexisting color sources. In the backward
evolution, on the other hand, the increment $\rmd Y=\epsilon$ is viewed as an increase in the
rapidity of the original quark; this looks `backwards', because the rapidity of the softest gluons is fixed.
Due to this boost, the quark can emit an early gluon, which becomes the new `first' gluon in the
evolution --- the one which is closest in rapidity to the original quark.

This `backward' perspective of the high-energy evolution is very convenient
in practice, since quite economical: at all the steps in the evolution, there is only source
of radiation --- the original quark. This viewpoint is underlying other related approaches,
such as the dipole picture by 
Mueller \cite{Mueller:1993rr} and the original constructions of the
Balitsky hierarchy \cite{Balitsky:1995ub} and of the BK equation 
\cite{Kovchegov:1999yj}. As we shall see, this viewpoint is also convenient 
for the inclusion of the collinear improvement and of the running coupling corrections
within the Langevin approach to the JIMWLK evolution.
 
As explained in the Introduction, the main virtue of this Langevin formulation  
is to allow for explicit numerical solutions
\cite{Rummukainen:2003ns,Lappi:2011ju,Dumitru:2011vk,Lappi:2012vw,Schlichting:2014ipa,
Schenke:2016ksl}. However, this formulation
is also useful for conceptual studies, like constructing the Balitsky equations
\cite{Balitsky:1995ub} for the evolution of products of Wilson line operators
(the elastic amplitudes for dilute projectiles). For the purpose of deriving differential
equations, it is sufficient to keep the terms that will matter up to order $\epsilon$ after 
performing the average over the noise. In this expansion, however, one should keep
in mind that the noise itself scales like $\nu\sim 1/\sqrt{\epsilon}$, as visible in \eqn{noise}.
Hence, the left and right infinitesimal rotations in \eqn{LangevinLO} must be expanded
up to {\em quadratic} order in their exponents. Moreover, the quadratic terms 
$\sim (\epsilon\nu)^2$ in this expansion can be already averaged over the noise, because to the
order of interest they cannot get multiplied by noise factors coming from 
other Wilson lines. After some simple algebra, one finds
\begin{align}\label{LangevExp}
U_{(n+1)\epsilon}^\dagger(\bx)
 \,=\, & U^\dagger_{n\epsilon}(\bx) 
   +\frac{\epsilon \alpha_s}{\pi^2}\int_{\bz}  {\mathcal K}_{\bx\bx\bz} \left(t^a U_{n\epsilon}^\dagger(\bx) \wt{U}_{n\epsilon}^{\dagger ab}(\bz)t^b - C_F U^\dagger_{n\epsilon}(\bx) \right)
  \nn
+\,&\rmi\,\frac{\epsilon\sqrt{\alpha_s}}{\pi}\int_{\bz} {\mathcal K}_{\bx\bz}^i \left( t^aU^\dagger_{n\epsilon}(\bx) - \tilde{U}^{\dagger ab}_{n\epsilon}(\bz)U^\dagger_{n\epsilon}(\bx)t^b \right)
\nu^{ia}_{n+1,\bz}+\order{\epsilon^{3/2}} \,,
\end{align}
where $C_F=(N_c^2-1)/{2N_c}$.
The term of $\order{\epsilon}$ in the first line has been obtained after averaging over the noise,
meaning that the gluon is both emitted and reabsorbed by the original quark at $\bx$.
This term contains two pieces: \texttt{(i)} a {\em real} term, 
involving the product of two Wilson lines, which describes a process where
the soft gluon has been emitted before the
collision and reabsorbed after it; \texttt{(ii)} a {\em virtual} term, involving the quark
Wilson line alone, which corresponds to processes where the gluon fluctuation has no
overlap in time with the scattering process. The term linear in
the noise in the second line of \eqn{LangevExp} is the {\em exchange} term,
which allows the gluon emitted by the quark at $\bx$ (either after, or prior to,
the collision) to be reabsorbed by some other Wilson line. \eqn{LangevExp} makes it clear
that at each step in the high-energy evolution, a new Wilson line is generated,
representing a soft gluon with a generic impact parameter.

\begin{figure}[t] \centerline{
\includegraphics[width=1.\textwidth]{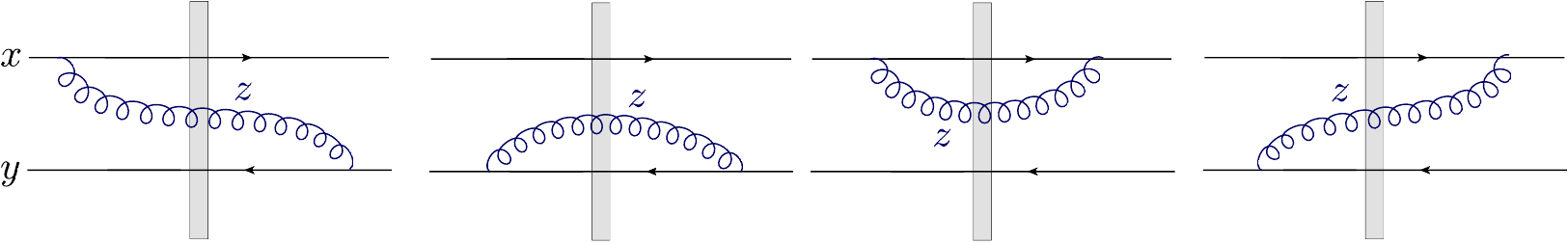}}
 \caption{\sl Diagrams illustrating one step in the BFKL evolution of a $q\bar q$ dipole via the emission
 of a soft `real' gluon (the gluon fluctuation lives at the time of scattering). The target is represented
 as a shockwave. In the (eight) corresponding
 `virtual' graphs, the gluon line is not crossing the shockwave.}
 \label{1gluon}
\end{figure}

As an application of \eqn{LangevExp}, let us use it to derive the first equation
in the Balitsky hierarchy \cite{Balitsky:1995ub}, that for the $S$--matrix of a quark-antiquark
color dipole. The dipole $S$--matrix is constructed as
\beq\label{Sdip}
S_Y(\bx,\by)\,\equiv\,\frac{1}{N_c}\left\langle \rmTr
 \big[U_{n\epsilon}^\dagger(\bx) U_{n\epsilon}(\by)\big] \right\rangle
 \eeq
where $Y=n\epsilon$, $\bx$ and $\by$ are the transverse coordinates of the quark and the
antiquark, and the average sign refers to both the noise average (over the
noise terms $\nu_1,\,...,\nu_n$ introduced by all the evolution steps) and the CGC average over
the target color field $A^-$ at $Y=0$ (the field in the initial Wilson line 
$U_{0}^\dagger(\bx)$). Using \eqn{LangevExp} together with a
similar equation for $U_{(n+1)\epsilon}(\by)$, taking the color trace and performing
the average over the noise $\nu_{n+1}$ associated with the last evolution step, one finds
\begin{align}\label{BKn}
 & \frac{1}{\epsilon} \left( \frac{1}{N_c} \rmTr \big[U_{(n+1)\epsilon}^\dagger(\bx) U_{(n+1)\epsilon}(\by)
\big] -\frac{1}{N_c}\rmTr \big[U_{n\epsilon}^\dagger(\bx) U_{n\epsilon}(\by)\big] \right) =
 \nn 
 & \qquad =\, \frac{\alpha_s}{\pi^2} \int_{\bz} {\mathcal M}_{\bx\by\bz} \left(\frac{1}{N_c}\rmTr 
\big[t^a U^\dagger_{n\epsilon}(\bx) t^b U_{n\epsilon}(\by)\big] \wt{U}_{n\epsilon}^{\dagger ab}(\bz) -C_F \frac{1}{N_c}\rmTr \big[U^\dagger_{n\epsilon}(\bx) U_{n\epsilon}(\by)\big] \right),
\end{align}
 where the {\em dipole kernel} (below, ${\mathcal K}_{\bx\by\bz}\equiv
 {\mathcal K}_{\bx\bz}^i {\mathcal K}_{\by\bz}^i$, cf. \eqn{Kdef})
\beq\label{Mdipole}
{\mathcal M}_{\bx\by\bz}\equiv {\mathcal K}_{\bx\bx\bz}+{\mathcal K}_{\by\by\bz}-2{\mathcal K}_{\bx\by\bz}=\frac{(\bx-\by)^2}{(\bx-\bz)^2(\bz-\by)^2}\,,
\eeq
has been generated by summing over the 4 possible topologies for the emission and the
absorption of the soft gluon, separately for `real' and `virtual' contributions (see Fig.~\ref{1gluon}).
For what follows, it is useful to notice that the dipole kernel can be factorized
as the product of two independent emissions:
\begin{align}\label{Mdipfact}
{\mathcal M}_{\bx\by\bz}=&\left({\mathcal K}_{\bx\bz}^i- {\mathcal K}_{\by\bz}^i\right)
\left({\mathcal K}_{\bx\bz}^i- {\mathcal K}_{\by\bz}^i\right)\nn
=&\int\frac{\rmd^2\bp}{2\pi}
\int\frac{\rmd^2\bq}{2\pi}\,\frac{\bp\cdot\bq}
 {p_\perp^2 q_\perp^2} \left(\rme^{\rmi \bp\cdot(\bx-\bz)}-\rme^{\rmi \bp\cdot(\by-\bz)}\right)
 \left(\rme^{-\rmi \bq\cdot(\bx-\bz)}-\rme^{-\rmi \bq\cdot(\by-\bz)}\right).
 \end{align}
That is, the soft gluon is first emitted, then reabsorbed, by either the quark or the antiquark. 
This factorized structure naturally emerges when deriving the Balitsky equations from the
JIMWLK Hamiltonian, which describes two independent, soft
gluon emissions from any of the preexisting partons.

As expected, the `real' term in the r.h.s. of \eqn{BKn} is the $S$--matrix of a 
quark-antiquark-gluon system. By using Fierz identities, this can be rewritten as 
a system of 2 quark-antiquark dipoles plus a single dipole contribution of $\order{1/N_c^2}$,
which precisely cancels against the respective contribution of the  `virtual' term in \eqn{BKn} (thus
replacing $C_F\to N_c/2$ in the coefficient of the latter). After also performing the relevant
averaging, one finds 
\begin{align}\label{BK}
 \frac{\del }{\del Y} \,S_Y(\bx,\by)=
 \frac{\abar}{2\pi}\, \int_{\bz}
 \mcal{M}_{\bx\by\bz}\Big[
 S_Y^{(2)}(\bx,\bz; \bz, \by)
 -S_Y(\bx,\by)\Big]\,,
 \end{align}
where $\abar\equiv\alpha_s N_c/\pi$ and $S_Y^{(2)}$ is the average $S$--matrix for a system
of two dipoles, defined as
\beq\label{S2dip}
S_Y^{(2)}(\bx_1,\by_1; \bx_2, \by_2)
\,\equiv\,\left\langle \frac{1}{N_c}\rmTr
 \big[U_{n\epsilon}^\dagger(\bx_1) U_{n\epsilon}(\by_1)\,
 \frac{1}{N_c}\rmTr
 \big[U_{n\epsilon}^\dagger(\bx_2) U_{n\epsilon}(\by_2)\big] \right\rangle\,.
 \eeq

\section{Time-ordering and collinear improvement}
\label{sec:TO}

In the physical discussion in the previous section, we have successively exposed the
viewpoints of target evolution and of projectile evolution, without making any
distinction between their physical consequences: the LO Langevin equation 
\eqref{LangevinLO} equivalently describes target evolution with decreasing $k^-$, or
projectile evolution with increasing $k^+$. Strictly speaking, the corresponding longitudinal
phase-spaces are not exactly the same, but their difference is irrelevant at LLA. However,
this difference becomes important beyond LO and to study this 
we need to more carefully specify the kinematics.

The rapidity phase-space for the target evolution is $Y^-\equiv \ln(P^-/p_0^-)$, where 
$p_0^-$ is the softest (in the sense of the lowest value for the longitudinal momentum $k^-$) 
gluon in the target wavefunction which matters for the scattering with the projectile. This value
$p_0^-$ is determined by the condition that the longitudinal wavelength 
$\Delta x^+ \simeq 1/p_0^-$ of this softest gluon be comparable with the lifetime 
$\Delta x^+ \simeq 2q^+/Q^2$ of the projectile. Here $q^+$ and $Q^2$ are the 
longitudinal momentum and respectively the virtuality of the projectile; e.g., for a dipole
projectile, like \eqref{Sdip}, one has $Q^2\sim 1/r^2$, with $r=|\bx-\by|$ the dipole transverse size.
This condition yields
\beq\label{p0}
\frac{1}{p_0^-}\,=\,\frac{2q^+}{Q^2}\ \Longrightarrow\,Y^-\,=\,\ln\frac{s}{Q^2}\,,\eeq
where $s=2P^-q^+$ is the energy squared in the center-of-mass frame. 
Similarly, the rapidity phase-space available for the high-energy evolution of the projectile is
$Y^+\equiv \ln(q^+/q_0^+)$, with $q_0^+$ determined by a condition analogous to \eqref{p0}:
\beq\label{q0}
\frac{1}{q_0^+}\,=\,\frac{2P^-}{Q^2_0}\ \Longrightarrow\,Y^+\,=\,\ln\frac{s}{Q^2_0}\,,\eeq
where we recall that $Q_0^2$ is the transverse resolution scale (generally, the saturation
momentum) of the un-evolved target.

The difference $Y^+-Y^-=\ln(Q^2/Q^2_0)\equiv \rho$ is positive in most interesting physical
situations, since in dilute-dense collisions the projectile looks generally `hard' ($Q^2\gg Q_0^2$)
on the resolution scale of the un-evolved target. In the LLA, one implicitly assumes
that this difference is negligible compared to the rapidity phase-space: $Y^+\simeq Y^-
\gg \rho$. When this happens, one can use either $Y^+$ or $Y^-$ as the `evolution time' $Y$,
and both target and projectile evolutions are correctly described (to LLA) by \eqn{LangevinLO},
or by the LO B-JIMWLK equations. However, there are important physical situations where
the rapidity shift $\delta Y=\rho$ cannot be neglected. First, $\rho$ is not necessarily
small compared to $Y$, not even in the high-energy kinematics; 
e.g., in the study of DIS at small Bjorken $x_{\rm Bj} \equiv Q^2/s$, 
the virtuality phase-space can be large enough for $\abar\rho \sim \order{1}$. Second, the
perturbative corrections to the high-energy evolution which are introduced by the transverse
phase-space are `anomalously large': they start at\footnote{When we indicate
the order of magnitude of the perturbative corrections throughout this paper, we mean
their magnitude relative to the leading order BFKL kernel, which is itself of $\order{\abar}$.}
$\order{\abar\rho^2}$ and not at $\order{\abar\rho}$, 
for reasons to be shortly reviewed. Accordingly, there is a
region in phase-space where $Y\gg \rho$, but such that both $\abar Y$ and
$\abar\rho^2$ are of order one, or larger. In that region, the evolution is still driven
by the rapidity, yet some of the `higher-order corrections'  enhanced are truly 
of  $\order{1}$ and must be resummed to all orders, for consistency.

At this level, the dissymmetry between target and projectile plays an important role. 
In the {\em collinear regime} where $\rho$ is positive
and large, say such that $\abar\rho^2\sim 1$, the projectile evolution in $k^+$ (or $Y^+$)
receives {\em double-logarithmic} corrections  $\sim (\abar\rho^2)^n$ to all orders, 
whereas the target evolution in $k^-$ (or $Y^-$) does {\em not} --- the respective
corrections involve only {\em single} collinear logarithms, of the form $(\abar\rho)^n$.
This dissymmetry can be attributed to the fact that the double collinear logarithms
are related to {\em time-ordering} in the high-energy evolution: the lifetime of the soft
daughter gluons should remain smaller than that of their faster parents. When $\rho$ is not
too large ($\abar\rho^2\ll 1$), this
condition is automatically satisfied, because of the strong ordering in longitudinal momenta.
But in the collinear regime, this condition might be violated by the LLA evolution of the projectile
in $Y^+$, while it is still satisfied by the LLA evolution of the target in $Y^-$. 

Recall indeed that the meaning of
`time' is different for the target and respectively the projectile.
For the left-moving target, the lifetime of a fluctuation is $\Delta x^-\simeq 2k^-/k_\perp^2$.
Successive emissions are strongly ordered in the longitudinal
momentum $k^-$, which decreases from the target towards the projectile.
In the collinear regime, they are typically ordered in $k_\perp$ as well, but in
the opposite direction: the transverse momentum increases when moving 
from the target towards the projectile (`soft-to-hard evolution'). 
This $k_\perp$--ordering is favored by the collinear phase-space: a sequence of $n$ gluon emissions
which are simultaneously ordered in $k^-$ ($P^-\gg k^-_1\gg k_2^-\cdots\gg k_n^-\gg p_0^-$)
and in $k_\perp$ ($Q_0 \ll k_{\perp 1}\ll k_{\perp 2}\cdots\ll k_{\perp n}\ll Q$) brings
in a large contribution, of order $(\abar Y\rho)^n$, which represents the dominant effect
of the LLA in this particular regime. (The approximation to LLA
which consists in keeping such contributions
alone is known as the {\em double-logarithmic approximation}, or DLA.) Clearly, for all such
typical configurations in the evolution of the target, the lifetimes are strongly 
ordered\footnote{This condition may be violated only for rare evolutions
where the transverse momenta are inversely ordered: $k_{\perp 2}\ll k_{\perp 1}$; however,
such evolutions are disfavored by the phase-space in the collinear regime and, moreover,
they are also cut off by gluon saturation in the target.}, as anticipated 
--- they decrease from the target
towards the projectile: $\Delta x^-_1\gg \Delta x^-_2\cdots \gg\Delta x^-_n$. 

For the right-moving projectile, on the other hand, the lifetime of a soft gluon is estimated as 
$\Delta x^+\simeq 2k^+/k_\perp^2$, where $k_\perp$ is strongly decreasing along the cascade
(`hard-to-soft evolution'), for the same reason as above: the strong bias from the collinear 
phase-space\footnote{This argument is not hindered by the saturation effects, which remain
small so long as $Q^2\gg k_\perp^2\gg Q_0^2$.}. 
Since $k^+$ and $k_\perp$ are simultaneously decreasing,
the lifetime of a daughter gluon can formally be larger than that of its parent parton.
This occurs indeed, in large regions of the phase-space, when the projectile evolution
in this collinear regime is computed according to LLA. But this situation is unphysical
and does {\em not} occur when the relevant Feynman graphs are
properly computed, beyond LLA: the phase-space regions where the time-ordering would be
reversed are automatically suppressed, by energy denominators 
(see \cite{Iancu:2015vea} for diagrammatic calculations which demonstrate this point).

The restriction to physical evolutions which respect the proper time-ordering has the effect
to reduce the rapidity phase-space for the projectile evolution, roughly from $Y^+$ to $Y^+-\rho$,
and thus introduces corrections to the LLA evolution which are enhanced by double collinear
logarithms --- that is, corrections of the type $(\abar\rho^2)^n$ \cite{Beuf:2014uia,Iancu:2015vea}.
To see this, consider the first step in the high-energy evolution of a dipole, namely the emission of
a soft gluon with $k^+\ll q^+$ by either the quark, or the antiquark, leg of the dipole
(recall Fig.~\ref{1gluon}). The lifetime condition implies
\beq\label{TO}
\frac{2q^+}{Q^2}\,>\,\frac{2k^+}{k_\perp^2}\ \Longrightarrow \ \Delta Y=\ln \frac{q^+}{k^+}
\,>\,\Delta \rho=\ln\frac{Q^2}{k_\perp^2}\,,\eeq
which is a genuine constraint if and only if $k_\perp^2\ll Q^2$ (since $\Delta Y > 0$
in any case). This must be true, in particular, for the softest gluon in this evolution,
that having $k^+=q_0^+$ and $k_\perp^2\sim Q^2_0$ (cf. \eqn{q0}); 
this implies $\ln(q^+/q^+_0) > \ln(Q^2/Q_0^2)$, or $Y \equiv Y^+ > \rho$. (From now on, we shall
denote $Y^+$ simply as $Y$, since we shall only study the projectile evolution.)
It is intuitively clear that one step in this constrained evolution brings an effect
$\abar(Y-\rho)\rho$, instead of the naive prediction ($\sim\abar Y\rho$) of
the unconstrained LLA (at DLA accuracy). Two successive such steps will produce an effect of order
\beq
[\abar(Y-\rho)\rho]^2\,=\,(\abar Y\rho)^2 - 2\abar\rho^2(\abar Y\rho) + (\abar\rho^2)^2.
\eeq
This can be formally interpreted as the cumulated effect of performing 2 unconstrained
steps in the DLA evolution (in $k^+$), plus an $\order{\abar\rho^2}$ correction to the 
emission kernel (the term linear in $Y$),
and a correction of $\order{(\abar\rho^2)^2}$ to the impact factor 
(the last term, independent of $Y$). 

Vice-versa, one can properly resum the corrections of the type $(\abar\rho^2)^n$ to all orders
by simply enforcing the time-ordering constraint within the equations describing the LLA evolution.
This has been already demonstrated at the level of the BK equation  \cite{Beuf:2014uia,Iancu:2015vea}
and in what follows we shall extend this strategy to the Langevin formulation of the JIMWLK
evolution and, hence, to the ensemble of the Balitsky equations at finite $N_c$.

But before we proceed, there is an important question that we would like to clarify: namely,
does the JIMWLK evolution require time-ordering in the first place ? That this is a legitimate question,
can be seen as follows: the JIMWLK equation has been originally constructed as {\em target}
evolution and we have just argued that, for the target evolution in $k^-$,
the proper time-ordering is automatically built-in at LLA, including in the collinear regime. The
situation of the JIMWLK equation is however special: albeit this was indeed constructed as
target evolution, the successive gluon emissions in this evolution were assumed to be
strongly ordered in $k^+$, and not in $k^-$ (the would-be natural evolution variable for
a left-mover). This can be checked by inspection of the manipulations in\footnote{See, e.g.,
the discussion in Sect. 3.4 of Ref.~\cite{Iancu:2000hn}; also, notice that in 
Refs.~\cite{Iancu:2000hn,Iancu:2001ad,Ferreiro:2001qy}, the dense target was chosen to be
a right-mover, hence the roles of $k^-$ and $k^+$ in the discussion there are reversed
as compared to the present discussion.} \cite{Iancu:2000hn,Iancu:2001ad,Ferreiro:2001qy}. 
This special choice was convenient for the inclusion of {\em gluon saturation}:
the soft gluons which are about to be emitted can multiply scatter off the strong background
field generated in the previous steps of the evolution; this field is localized in $x^+$,
hence its modes carry large components $k^-$, which will unavoidably alter the respective
momentum of the newly emitted gluon. Hence, $k^-$ is not a good `quantum number' for ordering
the evolution gluons. The $k^+$ component is better suited in that sense, since this is not modified
by rescattering (the background field is independent of $x^-$, by Lorentz time dilation).
These two components are related by the mass-shell condition (the evolution gluons are nearly
on-shell), $2k^+k^-=k_\perp^2$. Hence, {\em in the absence of any strong ordering in transverse
momenta}, the target evolution with decreasing $k^-$ is equivalent to its evolution
with increasing $k^+$. However, in the collinear regime of interest here, the transverse momenta
{\em are} strongly ordered (they increase simultaneously with $k^+$), so the time-ordering can
in fact be violated. To avoid that, one needs to amend the JIMWLK evolution by explicitly
enforcing time ordering. Another argument in that sense comes from the fact that
the JIMWLK evolution generates the Balitsky hierarchy, for which the necessity of time-ordering
is {\em a priori} clear from the perspective of projectile evolution.

\section{JIMWLK evolution with collinear improvement}
\label{sec:COLL}

Let us consider one step in the `backward' evolution of a dipole projectile and rewrite
the time-ordering constraint \eqref{TO} in terms of transverse sizes, rather than transverse
momenta. We recall that
$Q^2\simeq 1/r^2$, with $r=|\bx-\by|$ the transverse size of the parent dipole. 
Also, via the uncertainty principle, one can relate the transverse momentum of the 
soft gluon to the transverse sizes of the daughter dipoles:
$1/k_\perp\simeq{\rm min}(|\bx-\bz|, \,|\bz-\by|)$. The difference in size between 
the daughter dipoles is irrelevant in practice: the condition \eqref{TO} 
is non-trivial only when the daughter dipoles are sufficiently large, such that
$|\bx-\bz|\simeq |\bz-\by| \gg r$. Hence, for our purposes, one can replace
\eqn{TO} with
\beq\label{TOcoord}
q^+ r^2 \,>\,k^+ (\bx-\bz)^2\qquad\mbox{when}\qquad (\bx-\bz)^2\simeq
(\bz-\br)^2\gg r^2\,.\eeq
This constraint, together with the condition of strong ordering in longitudinal momenta,
$q^+\gg k^+\gg q_0^+$, can be conveniently summarized as two upper limits:
one on the transverse sizes $|\bx-\bz|\simeq |\bz-\by|$ of the (large) daughter dipoles, 
the other one on the corresponding rapidities.
Specifically, the condition $q^+ [r^2/(\bx-\bz)^2] > k^+\gg q^+_0$ implies 
\beq\label{theta}
Y\equiv \ln \frac{q^+}{q^+_0}\,>\,\rho^r_{\bx\bz} \equiv\ln\frac{(\bx-\bz)^2}{r^2}\,.\eeq
(This can also be viewed as a necessary condition for the existence of a soft emission with 
transverse size $|\bx-\bz|$, for a given rapidity separation $Y$ between the `valence' quark
and the target.) Also, for a given transverse size $|\bx-\bz| > r$ of the soft
gluon fluctuation, the maximal value of its rapidity $Y_k=\ln(k^+/q^+_0)$ that 
is allowed by time ordering\footnote{In the absence of time ordering, one has
$Y_{\rm max}=Y-\rmd Y$, which is the same as $Y_{\rm max}=Y$ to the accuracy of interest.}
reads, clearly,
\beq\label{shift}
Y_{\rm max}=Y- \Theta(|\bx-\bz|-r)\rho^r_{\bx\bz}\,,
\eeq
where the $\Theta$ function is intended to remind that this reduction in the rapidity
becomes effective only for sufficiently large daughter dipoles.

\subsection{Scale-dependent Wilson lines and the time-ordered Langevin equation}

It is rather straightforward to implement the kinematical constraints \eqref{theta} and \eqref{shift}
at the level of the BFKL or BK equation: the condition \eqref{theta} is inserted as a $\Theta$--function
multiplying the dipole kernel, whereas the rapidity shift \eqref{shift} is used to modify the
rapidity argument of the $S$--matrices for the daughter dipoles  \cite{Beuf:2014uia} (see also
\eqn{BKimprov} below). But when trying to apply a similar procedure to the Langevin equation
\eqref{LangevinLO}, one immediately faces the following difficulty: this equation contains 
no information about the overall size $r$ of the colorless projectile. (The transverse size of
the soft gluon fluctuation is properly encoded, as the distance $|\bx-\bz|$ between the gluon
and the quark.) Yet, whenever using \eqn{LangevinLO} in practice, one has in mind a 
well-identified projectile, whose transverse size (at tree-level) is {\em a priori} known. So, there is
no conceptual difficulty to further generalize\footnote{Recall that, as we argued in a previous 
section, within the projectile evolution picture, the `Wilson line' represents the scattering 
operator for the multi-partonic system generated by the high-energy evolution of a bare quark,
or gluon.} the notion of `Wilson line'  by associating to it a transverse size $r$ which 
represents the transverse spread of the colorless projectile to which that Wilson line belongs. For instance,
the two Wilson lines which enter the definition \eqref{Sdip} of the color dipole are promoted to
$U^\dagger_{n\epsilon}(\bx,r)$ and $U_{n\epsilon}(\by,r)$, respectively,
with $r=|\bx-\by|$. The additional
$r$--dependence is generated during the evolution, via the time-ordering constraints (see below).


A general projectile can be more complicated than just a dipole --- for instance, a quadrupole
is built with 4 partons (say, 2 quarks and 2 antiquarks)
already at tree-level, hence it can involve up to 6 different transverse sizes. 
In such a case, our present proposal for a generalized Wilson line applies only if
all the intrinsic transverse separations are comparable to each other (say, in such a
way that the logarithms $\ln({r_{ij}^2/r_{kl}^2})$ are of order one for all the pairs
$(ij)$ and $(kl)$ made with the valence partons). On the other hand, if the various
transverse sizes are very different from each other (at tree-level, once again), then
one cannot ignore the DGLAP-like evolution inside the wavefunction of the projectile,
so the problem goes beyond the scope of the high-energy evolution alone.

The fact that the proper definition of an {\em a priori} local operator at high energy
requires both longitudinal and transverse resolution scales is in line with 
what we know about the quantum evolution of the operators. These two scales control
the phase-space for the evolution, namely they fix the maximal longitudinal momentum 
$k^+_{\rm max}=q^+$ and the minimal size $r$ (or maximal transverse momentum $k_{\perp\rm max}^2
=1/r^2=Q^2$) of the quantum fluctuations permitted by our approximations.

The above considerations lead us to propose the following generalization of the Langevin equation
\eqref{LangevinLO}, which incorporates the kinematical constraints in Eqs.~\eqref{theta} and \eqref{shift}:
\beq\label{Langevincoll}
U_{(n+1)\epsilon}^\dagger(\bx,r) = \exp [\rmi\epsilon \alpha^L_{n+1}(\bx,r)] U^\dagger_{n\epsilon}(\bx,r)\exp [-\rmi\epsilon \alpha^R_{n+1}(\bx,r)],
\eeq
 where (compare to Eqs.~\eqref{left} and \eqref{right})
\begin{eqnarray}
\alpha^L_{n+1}(\bx,r)& =&\frac{1}{\pi}\int_{\bz} \sqrt{\alpha_s} \Theta\left(n\epsilon-\rho_{\bx\bz}^r\right){\mathcal K}_{\bx \bz}^i \nu_{n+1,\bz}^{ia}t^a, \label{leftcoll} \\
\alpha^R_{n+1}(\bx,r)&=&\frac{1}{\pi}\int_{\bz} \sqrt{\alpha_s} \Theta\left(n\epsilon-\rho_{\bx\bz}^r\right) {\mathcal K}_{\bx \bz}^i \nu_{n+1,\bz}^{ia} \wt{U}^{\dagger ab}_{n\epsilon- \Delta_{\bx\bz}^r}(\bz, R_{\bx\bz}^r)t^b. \label{rightcoll}
\end{eqnarray}
 We have introduced the notations
\beq\label{Delta}
\Delta_{\bx\bz}^r \equiv \Theta(|\bx-\bz|-r)\rho^r_{\bx\bz}\,,\qquad 
R^r_{\bx\bz}\equiv {\rm max}\{|\bx-\bz|,r\}.
\eeq
The noise correlator is the same as before, cf. \eqn{noise}. The $\Theta$--function inside
the integrands in Eqs.~(\ref{leftcoll}) and (\ref{rightcoll}) expresses the upper limit \eqref{theta}
on the transverse size of the gluon fluctuations, due to time-ordering. Also, the reduction in
the rapidity argument of the adjoint Wilson line, $n\epsilon\to n\epsilon- \Delta_{\bx\bz}^r$,
enforces the upper limit \eqn{shift} on the rapidity of the soft gluon. Because of this reduction,
the evolution described by Eqs.~(\ref{Langevincoll})--(\ref{rightcoll}) is non-local in $Y$.

 As anticipated,
the dependence upon the transverse resolution scale $r$ is generated by the evolution,
via the kinematical constraints in the Langevin equation. (The initial condition at $Y=0$ has no
such dependence: $U^\dagger_{0}(\bx,r)=U^\dagger_{0}(\bx)$. This can be selected,
e.g., according to the McLerran-Venugopalan model \cite{McLerran:1993ni}.) 
Note also the way how
this scale dependence gets updated when moving from the parent quark at $\bx$ to the
soft daughter gluon at $\bz$~: the corresponding scale $R^r_{\bx\bz}$ in the gluon Wilson
line is the largest among $r$ and $|\bx-\bz|$ since this is the characteristic distance 
for the transverse spread of the color charge during the lifetime of the fluctuation.

In writing Eqs.~(\ref{leftcoll}) and (\ref{rightcoll}), we moved the
QCD coupling $ \sqrt{\alpha_s}$ inside the integral over $\bz$, to anticipate the generalization
to a {\em running} coupling, which is made possible too by the presence of the reference
scale $r$. Specifically, the experience with running coupling effects in the context of
the DGLAP evolution (see e.g. the textbook  \cite{Kovchegov:2012mbw})
instructs us to re-interpret $\alpha_s$ in these equations as
\beq\label{run}
\alpha_s \,\equiv\, \alpha_s({\rm min}\{|\bx-\bz|,r\}).
\eeq
More precisely, the r.h.s. of \eqn{run} should be understood as the one-loop
running coupling,
 \begin{equation}\label{1Lrun}
 \alpha_s(k^2) =
 \frac{1}{b\ln\big[k^2/\Lambda_{\rm QCD}^2\big]},
 \end{equation}
 with $b=(11 N_c- 2 N_f)/12\pi$, evaluated at the virtuality scale $k^2\simeq 4/R^2$,
 with $R={\rm min}\{|\bx-\bz|,r\}$.
 
At this point, we would like to comment on a previous proposal \cite{Lappi:2012vw} 
for introducing a running coupling within the Langevin equation \eqref{LangevinLO}, but
without any additional scale dependence in the Wilson lines. In that case, the 
running of the coupling was encoded via a suitable modification of the noise correlator \eqref{noise}.
As a result, the QCD coupling was effectively evaluated at the scale $k^2=k_\perp^2$, with
$k_\perp$ the transverse momentum of the emitted gluon. Via the uncertainty principle,
this scale can be related to the transverse separation between the parent quark 
and the daughter gluon:
$k_\perp \sim 1/|\bx-\bz|$. This particular running coupling prescription is truly
correct so long as the emitted gluon is sufficiently hard, such that
$k_\perp^2\gtrsim Q^2$, or $|\bx-\bz|\lesssim r$. On the other hand, this prescription 
artificially enhances the relatively soft emissions with $|\bx-\bz|\gg r$,
by associating to them the large coupling\footnote{In \cite{Lappi:2012vw} it is argued
that the running of $\alpha_s$ with the transverse momentum $k_\perp$ of the emitted gluon 
should generate the correct, `smallest dipole', prescription at the level of the BK
equation (see the discussion around \eqn{rmin} below). However, this cannot be true,
since, whatever the kinematics, the gluon momentum $k_\perp$ is controlled by the transverse sizes, 
$|\bx-\bz|$ and $|\by-\bz|$,  of the daughter dipoles (and is insensitive to the size $r$ of the parent dipole). 
Specifically, the prescription in \cite{Lappi:2012vw} amounts to selecting $k_\perp=|\bp+\bq|/2$
as the argument of the coupling to be inserted in \eqn{Mdipfact} 
(see Eq.~(31) in \cite{Lappi:2012vw}). The integrals over $\bp$ and $\bq$ in \eqref{Mdipfact}
are independent from each other and they are limited in the ultraviolet by the daughter dipole
sizes, via the complex exponentials.
(This argument strictly holds 
for a fixed coupling, but the insertion of a running coupling
$\alpha_s(k_\perp^2)$ cannot change the convergence properties of these integrations.) 
In particular, for large daughter dipoles, $|\bx-\bz|\simeq
|\by-\bz|\gg r$, one necessarily has $k_\perp \sim 1/|\bx-\bz| \ll Q=1/r$.} 
$\alpha_s(|\bx-\bz|)$ instead of the
physical one $\alpha_s(r)$. This problem is remedied by our prescription in \eqn{run},
which looks natural in the context of the modified Langevin equation \eqref{Langevincoll},
but could not be implemented in the original equation \eqref{LangevinLO}, by lack of
explicit information about the projectile size $r$.

The proposal in Eqs.~(\ref{Langevincoll})--(\ref{run}) represents our main new result in this paper.
In what follows, we shall further motivate this proposal, in particular by showing that it leads to
an improved version of the BK equation which properly resums the double-logarithmic corrections
associated with time-ordering. Also, we shall extend this proposal to the resummation of an
important class of single-logarithmic corrections --- those which 
in perturbation theory start already at next-to-leading order.
 
 \subsection{The associated BK equation}

Let us first derive the generalization of the BK equation corresponding to \eqn{Langevincoll}.
To that aim, we proceed as explained in Sect.~\ref{sec:LO}, that is, we expand the
`left' and `right' color rotations in \eqn{Langevincoll} up to quadratic order in the noise and then
use (\ref{noise}) to average the quadratic terms. The analog of \eqn{LangevExp} reads
(with the shorthand notation $\Theta_{\bx\bz}\equiv\Theta(n\epsilon-\rho_{\bx\bz}^r)$)
\begin{align}\label{LangevExpColl}
U_{(n+1)\epsilon}^\dagger(\bx,r)
 \,=\, & U^\dagger_{n\epsilon}(\bx,r) 
   +\frac{\epsilon}{\pi^2}\int_{\bz}   \alpha_s\Theta_{\bx\bz}
   {\mathcal K}_{\bx\bx\bz} \left(t^a U_{n\epsilon}^\dagger(\bx,r) 
   \wt{U}_{n\epsilon-\Delta_{\bx\bz}^r}^{\dagger ab}(\bz,R_{\bx\bz}^r)
   t^b - C_F U^\dagger_{n\epsilon}(\bx) \right)
  \nn
+\,&\rmi\,\frac{\epsilon}{\pi}\int_{\bz} \sqrt{\alpha_s}
\Theta_{\bx\bz}{\mathcal K}_{\bx\bz}^i \left( t^aU^\dagger_{n\epsilon}(\bx,r) - \tilde{U}^{\dagger ab}_{n\epsilon-\Delta_{\bx\bz}^r}(\bz,R_{\bx\bz}^r)
U^\dagger_{n\epsilon}(\bx,r)t^b \right)
\nu^{ia}_{n+1,\bz}+\order{\epsilon^{3/2}} \,.
\end{align}

\comment{
\begin{eqnarray}
U_{(n+1)\epsilon}^\dagger(\bx,r) &\simeq& \left(1+\rmi\frac{\epsilon}{\pi}\int_{\bz}\sqrt{\alpha_s} 
\Theta_{\bx\bz}{\mathcal K}_{\bx\bz}^i\nu^{ia}_{n,\bz}T^a -\frac{\epsilon C_F }{2\pi^2}\int_{\bz}\alpha_s \Theta_{\bx\bz}{\mathcal K}_{\bx\bx\bz} \right) \, U^\dagger_{n\epsilon}(\bx,r) \nonumber \\ 
&&  \times \biggl( 1-\rmi\frac{\epsilon}{\pi} \int_{\bz}  \sqrt{\alpha_s}
\Theta_{\bx\bz} {\mathcal K}_{\bx \bz}^i \nu_{n,\bz}^{ia} \tilde{U}^{\dagger ab}_{n\epsilon-\Delta_{\bx\bz}^r}(\bz,R_{\bx\bz}^r )T^b  -\frac{\epsilon C_F}{2\pi^2}\int_{\bz} \alpha_s
\Theta_{\bx\bz} {\mathcal K}_{\bx\bx\bz} \biggr) 
\end{eqnarray}
where we have used the shorthand notation $\Theta_{\bx\bz}=\Theta(Y-\rho_{\bx\bz}^r)$.
}

Using this together with the corresponding Langevin equation for the antiquark
Wilson line $U_{n\epsilon}(\by,r)$ and performing the average over $\nu^{ia}_{n+1}$, 
one finds the following evolution equation for the dipole $S$--matrix (with $Y=n\epsilon$ and $r=|\bx-\by|$):
\begin{eqnarray}
&&\frac{\rmd}{\rmd Y} \frac{1}{N_c}\rmTr [U^\dagger_{Y}(\bx,r) U_{Y}(\by,r)] \nn 
&& = \frac{N_c}{2\pi^2}\int_{\bz} \biggl\{ 
 \left(\alpha_s \Theta_{\bx\bz} {\mathcal K}_{\bx\bx\bz}-\sqrt{\alpha_s\alpha_s} \Theta_{\bx\bz}\Theta_{\by\bz} {\mathcal K}_{\bx\by\bz}\right)  
\nn 
&&
 \qquad \qquad \qquad \times \frac{1}{N_c}\rmTr [U^\dagger_{Y}(\bx,r) U_{Y-\Delta_{\bx\bz}^r}(\bz,R_{\bx\bz}^r)] \frac{1}{N_c} \rmTr [U^\dagger_{Y-\Delta_{\bx\bz}^r}(\bz,R_{\bx\bz}^r )U_{Y}(\by,r)]
 \nn 
&&\qquad\qquad +\left(\alpha_s \Theta_{\by\bz} {\mathcal K}_{\by\by\bz}-\sqrt{\alpha_s\alpha_s} \Theta_{\bx\bz}\Theta_{\by\bz} {\mathcal K}_{\bx\by\bz}\right)
\nn
&&\qquad\qquad
 \qquad \times \frac{1}{N_c}\rmTr [U^\dagger_{Y}(\bx,r) U_{Y -\Delta_{\by\bz}^r}(\bz,R_{\by\bz}^r)] \frac{1}{N_c}\rmTr [U^\dagger_{Y-\Delta_{\by\bz}^r}(\bz,R_{\by\bz}^r)U_{Y}(\by,r)]  
\nn
&& \qquad
-\left( \alpha_s \Theta_{\bx\bz}{\mathcal K}_{\bx\bx\bz}+\alpha_s\Theta_{\by\bz}{\mathcal K}_{\by\by\bz }-\sqrt{\alpha_s\alpha_s}\Theta_{\bx\bz}\Theta_{\by\bz}{\mathcal K}_{\bx\by\bz} \right) \frac{1}{N_c} \rmTr [U_{Y}^\dagger(\bx,r) U_{Y}(\by,r)] \nn
&& \qquad
+\sqrt{\alpha_s\alpha_s}\Theta_{\bx\bz}\Theta_{\by\bz}  {\mathcal K}_{\bx\by\bz} \frac{1}{N_c} \rmTr\left[ U_{Y-\Delta_{\bx\bz}^r}^\dagger(\bz,R_{\bx\bz}^r)   U_{Y-\Delta_{\by\bz}^r}(\bz,R_{\by\bz}^r)\right] 
\nn 
&& \qquad
 \qquad \qquad \qquad \times \frac{1}{N_c}\rmTr \Bigl[U_{Y}^\dagger (\bx,r)U_{Y-\Delta_{\bx\bz}^r}(\bz,R_{\bx\bz}^r) U^\dagger_{Y-\Delta_{\by\bz}^r} (\bz,R_{\by\bz}^r)U_{Y}(\by,r)\Bigr]  \biggr\}. \label{BKcoll}
\end{eqnarray}
The averaging over the noise terms $\nu_1,\,...,\nu_n$ from the previous evolution
steps and over the
initial Wilson line $U_0$ is kept implicit: at this stage, this averaging is not needed and \eqn{BKcoll} 
can be as well understood at operatorial level.
The argument of the running coupling has been suppressed
to keep the notations simple, but can be easily restored from the accompanying 
$\Theta$--function, i.e., $\sqrt{\alpha_s}\Theta_{\bx\bz}\equiv
\sqrt{\alpha_s({\rm min}\{|\bx-\bz|,r\})}\Theta_{\bx\bz}$. Also, we have used Fierz identities 
to rewrite the various $S$--matrices in terms of fundamental Wilson lines alone.

The first two terms in the r.h.s. of  
\eqn{BKcoll} refer to `real' emissions, that is, to gluons which cross the shockwave and 
can interact with it: the first term describes the case where the soft gluon is emitted
prior to the collision by the quark, whereas the second term similarly refers to an
emission by the antiquark. The two other terms in \eqn{BKcoll} refer to `virtual' emissions. 
The first one, involving the original dipole $(\bx,\,\by)$, has a familiar structure. 
But the second one involves new color structures: a color {\em quadrupole}, built with four
Wilson lines, together with an unusual dipole whose both legs lie at $\bz$.
This last term describes the situation where the gluon is exchanged
between the quark and  the antiquark before the collision.

\eqn{BKcoll} looks considerably more complicated than the usual BK equation \eqref{BK}, 
but this is not a problem in practice, since we do not really need to solve this equation
--- all numerical efforts should directly focus on the Langevin equation \eqref{Langevincoll}. Here,
we shall use \eqn{BKcoll} merely to illustrate the effects of the time-ordering. Besides, we shall
argue that, to the accuracy of interest, this equation can be further simplified.


Namely, the complications apparent in \eqn{BKcoll} are largely due to the
differences between the kinematical constraints associated with the quark ($\bx$) and,
respectively, the antiquark ($\by$) --- e.g., the fact that, in general, the $\Theta$--functions
$\Theta_{\bx\bz}$ and $\Theta_{\by\bz}$ are different from each other. Such differences 
prevented us from reconstructing the dipole kernel according to \eqn{Mdipole}; they
also explain the appearance
of the new color structures in the last term in  \eqn{BKcoll}. But in reality,
these differences go beyond the accuracy of interest, as anticipated above
\eqn{TOcoord} and it will be further explained.

Notice first that the `ultraviolet' structure of \eqn{BKcoll}
--- i.e., the behavior of its integrand in the limit where the daughter gluon 
lies very close to its parent quark ($|\bx-\bz|\ll r$) or antiquark ($|\bz-\by|\ll r$) --- is not
affected by time-ordering. Indeed, in any of these limits, the kinematical constraints become irrelevant
and \eqn{BKcoll} reduces to the original equation \eqref{BK}. (For instance,
when $|\bx-\bz|\ll r$, one has $\rho^r_{\bx\bz} < 0$ and $\rho^r_{\bz\by}\simeq 0$, hence
$\Theta_{\bx\bz}=\Theta_{\by\bz}=1$, $\Delta_{\bx\bz}^r=\Delta_{\bz\by}^r=0$,
 and $R_{\bx\bz}^r=R_{\bz\by}^r=r$.) This was to be expected, 
 since the time-ordering can only affect the gluon emissions with
relatively soft transverse momenta. This is moreover important, as it ensures that the 
short-distance divergences of the emission kernels at  $|\bx-\bz|\to 0$ and $|\bz-\by|\to 0$ 
are harmless, since the `real' and `virtual' terms mutually cancel in these limits
--- as in the standard BK equation \eqref{BK}.

Consider now the opposite limit, where the daughter dipoles are relatively large,
$|\bx-\bz|\simeq |\bz-\by| \gg r$. In that case, the time-ordering is clearly important, but the
associated kinematical constraints look identical for the emissions by the quark and the antiquark,
respectively. This, together with the unitarity of the Wilson lines, implies that the quadrupole 
piece which appears in the last term in \eqn{BKcoll} reduces to the original dipole $(\bx,\,\by)$, 
whereas the dipole piece within the same term reduces to unity. Therefore, it becomes possible 
to combine the various terms in the r.h.s. of \eqn{BKcoll} in such a way to reconstruct the 
dipole kernel according to \eqn{Mdipole}. Then the structure of the `improved' equation becomes 
quite similar to that of the original equation \eqref{BK}, namely,
\begin{align}\label{BKimprov}
 \frac{\del }{\del Y} \,\hat S_Y(\bx,\by)=
 \frac{ \alpha_s(r) N_c}{2\pi^2}\, \int_{\bz}\Theta_{\bx\bz} 
 \mcal{M}_{\bx\by\bz}\Big[
\hat S_{Y-\Delta_{\bx\bz}^r}(\bx,\bz)\hat S_{Y-\Delta_{\bx\bz}^r}(\bz,\by) -\hat S_Y(\bx,\by)\Big]\,,
 \end{align}
where we identified the dipole $S$-matrices which enter the `real' term according to\begin{eqnarray}
\frac{1}{N_c}\rmTr [U^\dagger_{Y}(\bx,r) U_{Y-\Delta_{\bx\bz}^r}(\bz,R^r_{\bx\bz})] = \hat S_{Y-\Delta_{\bx\bz}^r}(\bx,\bz)\,.  \label{Sshift}
\end{eqnarray}
Notice that the two legs of this dipole live at different rapidities and it is the {\em softest} leg
(here, the antiquark piece of the gluon at $\bz$) 
which fixes the rapidity argument of the overall $S$--matrix.
This is appropriate since the rapidity phase-space which remains available for the evolution
after the emission of the first gluon is not $Y$, but $Y-\Delta_{\bx\bz}^r$. The `hat' symbol on the
$S$--matrix is intended to emphasize that this is an {\em operator}, like the Wilson lines themselves.
In particular, there is no factorization assumption in \eqn{BKimprov}: this holds
for generic $N_c$. 

\eqn{BKimprov} has been written for the situation where $|\bx-\bz|\simeq |\bz-\by| \gg r$ and
hence one can identify $\Theta_{\bx\bz}= \Theta_{\by\bz}$ and
$\Delta_{\bx\bz}^r= \Delta_{\bz\by}^r= \ln[{(\bx-\bz)^2}/{r^2}]$.
But as a matter of facts, this equation can be extended to all
physical regimes, as a replacement for the more complicated equation \eqref{BKcoll}.
To that aim, it suffices to replace, in \eqn{BKimprov}, 
$\Theta_{\bx\bz}\to \Theta_{\bx\bz} \Theta_{\bz\by}= \Theta(Y-\rho_{\bx\by\bz})$, 
 $\Delta_{\bx\bz}^r\to \Delta_{\bx\by\bz}$, and
$\alpha_s(r)\to \alpha_s(r_{\rm min})$,
where 
\beq
 \rho_{\bx\by\bz}\equiv {\rm max}\big(\rho_{\bx\bz}^r,
\rho_{\by\bz}^r\big) \,,\qquad
\Delta_{\bx\by\bz}\equiv \Theta(\rho_{\bx\by\bz})\,\rho_{\bx\by\bz}\,,\eeq
and
\beq\label{rmin} 
r_{\rm min} \equiv \min\big\{|\bx \minus\by|,|\bx \minus\bz|,|\by \minus\bz|\big\}\,.
\eeq
Indeed, after these replacements, Eqs.~\eqref{BKimprov} and \eqref{BKcoll} coincide
with each other for both very small, and very large, daughter dipoles, whereas the  
differences between them which occur when the three dipoles are commensurable to each other
($|\bx-\bz|\sim |\bz-\by| \sim r$) are irrelevant to the accuracy of interest
(indeed, in that case, $\rho^r_{\bx\bz}$ and $\rho^r_{\bz\by}$ are both of $\order{1}$ and hence
negligible compared to $Y$).

The generalization of \eqn{BKimprov} that we have just described is 
equivalent (to the accuracy of 
interest, once again) to the collinearly-improved version of the BK equation proposed 
in \cite{Beuf:2014uia} as a tool to resum the double-logarithmic corrections
$\sim (\abar\rho^2)^n$ to all orders. Furthermore, as shown by the diagrammatic analysis in 
\cite{Iancu:2015vea}, this equation follows directly from the QCD Feynman graphs provided
one uses a convenient organization of the perturbation theory --- namely, the light-cone
(or time-dependent, with `time' = $x^+$) perturbation theory in the projectile light-cone
gauge $A^+=0$.

\subsection{Partially resumming single logarithms}

In this subsection we shall present a further refinement of the Langevin equation,
which refers to a partial resummation of the radiative corrections enhanced by {\em single}
collinear logarithms --- that is, the corrections of the type $(\abar\rho)^n$,
which represent the interplay between the BFKL and the DGLAP evolutions.
Namely, as shown in Ref.~\cite{Iancu:2015joa}, it is relatively easy to resum a particular subset of
such corrections, which fully includes the respective piece at NLO (that is, the correction of
order $\abar\rho$ to the BFKL kernel), together with a part of the higher order terms. Albeit
incomplete, this resummation is still useful in practice, in that it allows one to keep under
control all the NLO BFKL corrections which are
amplified by (double or single) transverse logarithms. 
In other terms, the final version of the Langevin equation to be presented below
is perturbatively correct up to pure $\order{\abar}$ corrections to the BFKL kernel.

Diagrammatically, the single-logarithmic corrections of interest correspond to Feynman graphs in
which one `BFKL emission' (the emission of a gluon with small longitudinal momentum
fraction) is accompanied by an arbitrary number $n\ge 1$ of `DGLAP emissions'
(quarks or gluons with longitudinal momentum fractions of $\order{1}$ and whose
transverse momenta are strongly ordered). The transverse momenta can be either
decreasing (`collinear resummation'), or increasing (`anticollinear'), along the cascade.
In the context of the linear, BFKL, evolution, one was able to fully resum such corrections,
by working in a double Mellin representation where the DGLAP-like corrections exponentiate
\cite{Kwiecinski:1997ee,Salam:1998tj,Ciafaloni:1999yw,Ciafaloni:2003rd,Vera:2005jt}.
Here, we shall only resum a subset of these corrections which
exponentiates already in transverse coordinate (or momentum) space.
This partial resummation involves the first moment of a particular
linear combination of the DGLAP splitting functions, namely,
\begin{equation}
    \label{pomega}
 	\int_0^1 \dif z\,  
 	\left[ \left(P_{\rm gg}(z)-\frac{2N_c}{z}\right) + \frac{\CF}{\Nc}\, P_{\rm qg}(z) \right]
	= - 2N_c\left[\frac{11}{12} + \frac{\Nf}{6\Nc^3}\right]
 	\equiv - 2N_c A_1.
 \end{equation} 
$P_{\rm gg}(z)$ and $P_{\rm qg}(z)$ are the gluon-to-gluon and, respectively,
gluon-to-quark leading-order DGLAP splitting functions, in the conventions
of  \cite{Kovchegov:2012mbw} (see Eqs.~(2.98)--(2.99) in that textbook).
With these conventions, $P_{\rm gg}(z)$ has a singular piece $2N_c/z$ at $z\to 0$, which has been
subtracted in \eqn{pomega}, as this would describe a BFKL emission. 
The particular linear combination of splitting functions which occurs in the l.h.s. of \eqn{pomega}
reflects the fact that the gluon and quark distribution functions are coupled under the DGLAP evolution.
For the present purposes, we have selected the eigenvalue of the $2\times 2$ matrix-valued 
anomalous dimension\footnote{We recall that the DGLAP anomalous dimension 
$\mathcal{P}_{ij}(\omega)$ associated with a generic partonic splitting function $j\to i$
is computed as  $\mathcal{P}_{ij}(\omega) = \int_0^1 \rmd z\,z^{\omega} P_{ij}(z)$.
The first moment in \eqn{pomega} singles out the first non-singular term in
the small-$\omega$ expansion of the relevant eigenvalue:  $ \mathcal{P}(\omega) = 
2N_c[{1}/{\omega} - A_1  + \mathcal{O}(\omega)]$.}
 which controls the growth of the dipole amplitude with increasing energy (see e.g.
 \cite{Ciafaloni:1999yw} for details).
Notice also that we have defined
the `anomalous dimension' $A_1$ to be positive definite.

We shall first describe the resummation of the single-logarithmic corrections at the level of the
BK equation.  For a fixed coupling,
the corresponding generalization of \eqn{BKimprov} reads (as compared to
\eqn{BKimprov}, we also extend the kinematical constraints as discussed around \eqn{rmin})
\begin{align}\label{BKfixed}
 \frac{\del }{\del Y} \,\hat S_Y(\bx,\by)=&\,
 \frac{ \abar}{2\pi} \int_{\bz} \Theta(Y-\rho_{\bx\by\bz})\,
  \mcal{M}_{\bx\by\bz} \left[\frac{r^2}{r_<^2}\right]^{\pm \abar A_1}
  \nn &\qquad\qquad\qquad\qquad\qquad\times
\Big[
\hat S_{Y-\Delta_{\bx\by\bz}}(\bx,\bz)\hat S_{Y-\Delta_{\bx\by\bz}}(\bz,\by) -\hat S_Y(\bx,\by)\Big],
 \end{align}
with $r_<^2\equiv \min\{(\bx \minus \bz)^2,(\by \minus \bz)^2\}$.
The factor within the square brackets in the r.h.s., whose exponent features
$A_1$, is the result of resumming the single logarithms: the positive sign in the
exponent is taken when  $r < r_<$ 
and the negative sign otherwise. Recalling that $A_1$ is positive, we see that this
resummation is always slowing down the evolution. \eqn{BKfixed} differs from the
collinearly-improved version of the BK equation proposed in \cite{Iancu:2015joa} in
that the effects of the time-ordering for the collinear emissions are encoded via
kinematical constraints, and not via a resummation of the kernel. Yet, these two equations
(\eqn{BKfixed} above and Eq.~(9) in Ref.~\cite{Iancu:2015joa}) are equivalent to
the accuracy of interest.

The generalization of \eqn{BKfixed} to a one-loop running coupling reads (cf. \eqn{run})
\begin{align}\label{BKfin}
 \frac{\del }{\del Y} \,\hat S_Y(\bx,\by)=&\,
 \frac{1}{2\pi} \int_{\bz} \abar(r_{\rm min}) \Theta(Y-\rho_{\bx\by\bz})\,
  \mcal{M}_{\bx\by\bz}\left[\frac{\abar(r)}{\abar(r_<)}\right]^{\pm {A_1}/{\bar b}}\nn
   &\qquad\qquad\qquad\qquad\qquad\times
\Big[
\hat S_{Y-\Delta_{\bx\by\bz}}(\bx,\bz)\hat S_{Y-\Delta_{\bx\by\bz}}(\bz,\by) -\hat S_Y(\bx,\by)\Big],
 \end{align}
where ${\bar b}=\pi b/N_c$
and the positive (negative) sign in the exponent applies when  $r< r_<$ (respectively, $r> r_<$).
The running coupling version
of the factor resumming the single logarithms has been obtained via the following argument
(which is familiar in the context of the DGLAP evolution; see e.g.  \cite{Ciafaloni:2003rd} for
a similar discussion). Given two widely separated transverse momentum (or virtuality) 
scales $Q^2\gg Q_0^2$,  the resummation of the DGLAP logs $[\alpha_s\ln(Q^2/Q_0^2)]^n$ 
which is relevant for our purposes at {\em fixed} coupling reads (compare to  \eqn{BKfixed})
\beq
\left(\frac{Q_0^2}{Q^2}\right)^{\abar A_1}=\,\rme^{-\abar A_1\ln\frac{Q^2}{Q_0^2}}\,.\eeq
With a one-loop running coupling, this is replaced by
\beq
\exp\left\{-\frac{A_1}{\bar b}\int_{\rho_0}^\rho\frac{\rmd \rho'}{\rho'}\right\}=
\exp\left\{-\frac{A_1}{\bar b}\ln\frac{\rho}{\rho_0}\right\}=\left[\frac{\abar(Q^2)}{\abar(Q^2_0)}
\right]^{A_1/{\bar b}},
\eeq
as shown in \eqn{BKfin}. Within the above equation, $\rho\equiv \ln(Q^2/\Lambda^2)$
and similarly $\rho_0\equiv \ln(Q^2_0/\Lambda^2)$.

We are now prepared to present the extension of the Langevin equation 
which resums single-logarithms as well: this has the same general structure as shown in
\eqn{Langevincoll}, but with the arguments of the infinitesimal, `left' and `right', rotations
modified to (for the case of a running coupling)
\begin{eqnarray}
\alpha^L_{n+1}(\bx,r)& =&\frac{1}{\pi}\int_{\bz} \sqrt{\alpha_s} \Theta\left(n\epsilon-\rho_{\bx\bz}^r\right)
\left[\frac{\abar(r)}{\abar(|\bx-\bz|)}\right]^{\pm {A_1}/{2{\bar b}}}
{\mathcal K}_{\bx \bz}^i \nu_{n+1,\bz}^{ia}t^a, \label{leftfin} \\
\alpha^R_{n+1}(\bx,r)&=&\frac{1}{\pi}\int_{\bz} \sqrt{\alpha_s} \Theta\left(n\epsilon-\rho_{\bx\bz}^r\right) \left[\frac{\abar(r)}{\abar(|\bx-\bz|)}\right]^{\pm {A_1}/{2{\bar b}}}
{\mathcal K}_{\bx \bz}^i \nu_{n+1,\bz}^{ia} \wt{U}^{\dagger ab}_{n\epsilon- \Delta_{\bx\bz}^r}(\bz, R_{\bx\bz}^r)t^b, \label{rightfin}
\end{eqnarray}
where, as before, the factor of $\alpha_s$ under the square root is understood
as $\alpha_s\equiv \alpha_s({\rm min}\{|\bx-\bz|,r\})$; also, the positive sign in the
exponent is taken when  $r< |\bx \minus \bz|$ and the negative sign when  $r> |\bx \minus \bz|$.

\comment{Hence,
the new factor in the structure of Eqs.~\eqref{leftfin}--\eqref{rightfin} is the same as
\beq\label{SL}
\left[\frac{r^2}{(\bx \minus \bz)^2}\right]^{\pm \frac{\abar A_1}{2}}
=\,\exp\left\{-\frac{\abar A_1}{2}|\rho_{\bx\bz}^r|\right\}.
\eeq}

It is quite obvious that the version of the BK equation which corresponds to this new
Langevin equation can be immediately obtained by multiplying all the Weizs\"{a}cker-Williams kernels
in \eqn{BKcoll} by the new factor appearing
in Eqs.~\eqref{leftfin}--\eqref{rightfin} (or the analog factor with $\bx\to\by$).
The ensuing equation looks considerably more complicated than \eqn{BKfin} above,
yet they are equivalent at the accuracy of the present approximations:
In the collinear regime where the daughter dipoles are large, $|\bx-\bz|\simeq |\bz-\by| \gg r$, 
the additional factor associated with single logarithms
becomes the same for emissions by the quark and respectively the antiquark, hence
it appears as a factor multiplying the dipole kernel, such as in \eqn{BKfin}. In the opposite,
anticollinear, regime, where one of the daughter dipoles is very small, 
e.g. $|\bx-\bz|\ll r\simeq |\bz-\by|$, then both \eqn{BKfin} and \eqn{BKcoll}  are dominated by
processes where the soft gluonis emitted and reabsorbed by the nearby quark at $\bx$; 
for instance, $\mcal{M}_{\bx\by\bz}\simeq {\mathcal K}_{\bx\bx\bz} = 1/(\bx-\bz)^2$.
For these processes, the Langevin equation obviously produces the same `single-logarithm'
factor as that visible in \eqn{BKfin}, namely $[\abar(|\bx-\bz|)/\abar(r)]^{A_1/{\bar b}}$. The mismatch between
the respective factors for the other processes (which also involve the antiquark at $\by$)
is irrelevant, since those processes are anyway negligible in the anticollinear limit.

\section{Summary and conclusions}
\label{sec:conc}

In this paper, we have constructed a collinearly-improved version of the leading order JIMWLK
evolution which allows for a complete resummation of the large radiative corrections enhanced
by double collinear logarithms together with partial resummations of the single collinear logarithms
and of the running coupling corrections. In particular, all the (double and single) transverse logarithms 
which enter the next-to-leading corrections to the high-energy evolution of Wilson lines are correctly
included in our scheme and at the same time kept under control via all-order resummations. Hence
our results provide an extension to the full Balitsky-JIMWLK hierarchy --- meaning to generic
values for the number of colors $N_c$ and to Wilson-line correlators more complicated than
just a dipole --- of the collinear improvement for the BK equation previously presented in 
\cite{Beuf:2014uia,Iancu:2015vea,Iancu:2015joa}. Most importantly, our strategy for collinear improvement
is implemented directly at the level of the Langevin formulation of the LO JIMWLK evolution,
which is indeed the most useful formulation for numerical calculations. The same strategy
can be applied to the more general Langevin equations which describe the high-energy
evolution of multi-particle production with rapidity correlations in dilute-dense collisions 
\cite{Iancu:2013uva}.

An essential ingredient of our method is a generalization of the Wilson line operators, which are
now allowed to depend upon an additional transverse scale --- the typical 
transverse size of the color-neutral projectile to which belongs the Wilson line under consideration.
This dependence is merely logarithmic --- it is responsible for the resummation of the collinear 
logarithms alluded to above ---, hence our strategy can also be used for multi-parton projectiles, 
which involve several such scales, so long as these scales are not too widely separated from each other.
While natural for the purposes of the quantum evolution with the transverse resolution, this
additional scale dependence has nevertheless the drawback to complicate the numerical calculations.
So it remains as a challenge to provide numerical implementations for our proposal.

Our strategy for resumming the double logarithmic corrections is in the spirit of
Ref.~\cite{Beuf:2014uia}, in that the time-ordering has been implemented via kinematical constraints
on the evolution phase-space. This led to evolution equations which are {\em non-local} in rapidity
--- a feature which should not represent a major impediment in practice (since easily to accommodate
within the iterative procedure that is generally used for solving Langevin equations). But this rises
an interesting conceptual issue: being non-local in $Y$, our `improved' Langevin equations cannot 
be generated by a suitable generalization of the JIMWLK Hamiltonian. 
In turn, this rises the question about the possibility to write down a
collinearly-improved version of the leading order JIMWLK Hamiltonian. At this point, this question
may look a bit academic, given that we already know how to improve the corresponding 
Langevin formulation. Yet, this is interesting as it might shed more light on the structure of the
high-energy evolution beyond leading order.
As we now explain, this question can be addressed at two different levels 
--- a `weak' level and a `strong' one--- and the respective answers are not 
the same.  

The `weak' level refers to the dipolar version of the LO JIMWLK Hamiltonian 
\cite{Hatta:2005as}, as  obtained by replacing 
$\mcal{K}_{\bx\by\bz}\to \mcal{M}_{\bx\by\bz}$ within the original JIMWLK Hamiltonian
\cite{Weigert:2000gi,Iancu:2001ad,Ferreiro:2001qy}. This is `weak' in the sense that
it does not admit a Langevin formulation, hence it cannot be used for numerical calculations.
Yet, this is useful for formal studies, e.g. as an efficient tool for deriving the equations
in the Balitsky hierarchy. A collinearly-improved version
of this `dipole' Hamiltonian can be easily deduced  from the results 
in \cite{Iancu:2015vea,Iancu:2015joa}. Namely, in those papers, one has constructed
a colllinearly-improved BK equation, which is {\em local} in rapidity and
where the resummation of the double collinear logarithms is performed directly 
at the level of the kernel. That is, the effects of the kinematical constraints appearing 
in Eqs.~\eqref{BKimprov} or \eqref{BKfixed} have been equivalently 
replaced by a change in the kernel, 
$\mcal{M}_{\bx\by\bz}\to \mcal{M}_{\bx\by\bz}\mathcal{K}_{\sdla}$,
with $\mathcal{K}_{\sdla}$ resumming powers of $\abar\rho^2$ to all orders
(compare the present \eqn{BKfixed} with Eq.~(9) in \cite{Iancu:2015joa}).
The correspondingly improved version of the `dipole' JIMWLK Hamiltonian can be
obtained via a similar replacement. Yet, this suffers from the drawback that we just
mentioned: it does not lend itself to a Langevin formulation.

The `strong' level refers to a Fokker-Planck--like Hamiltonian which has a factorized
structure and thus allows for a Langevin formulation
--- such as the leading-order JIMWLK Hamiltonian (see the discussion in \cite{Blaizot:2002xy}).
In our opinion, this factorized structure is not consistent with the collinear improvement, 
more precisely, with the constraint of time-ordering. There are two ways to see that. 
On one hand, the improved Langevin equations that we have here obtained 
are non-local in rapidity and cannot be rewritten in a local 
form, via manipulations similar to those in Refs.~ \cite{Iancu:2015vea,Iancu:2015joa}.
(Indeed, the strategy in  \cite{Iancu:2015vea,Iancu:2015joa} has crucially relied on the
possibility to construct exact analytic solutions for the evolution equation with time
ordering, in the double logarithmic approximation. This is clearly not
the case for the time-ordered Langevin equations, due to their stochastic nature.)
On the other hand, the collinearly-improved BK equation obtained in 
\cite{Iancu:2015vea,Iancu:2015joa}, which {\em is} local in $Y$, cannot be
generated by a Hamiltonian of a Fokker-Plank type, because the new factor
$\mathcal{K}_{\sdla}$ in the kernel cannot be factorized as the product of two 
emissions, in contrast to the LO dipole kernel (recall \eqn{Mdipfact}).

These considerations suggest that the probabilistic picture underlying the CGC effective 
theory \cite{Iancu:2002xk,Iancu:2003xm,Gelis:2010nm},
that is heavily relying on the Fokker-Planck nature of the evolution Hamiltonian, may not survive
beyond leading order. This is not necessarily a surprise: a similar situation occurs in the more 
familiar context of the collinear factorization, where the probabilistic interpretation of the structure
functions ceases to be valid beyond leading order, notably due to the scheme dependence
of the higher order corrections. The fact that the largest radiative corrections to the high-energy
evolution can nevertheless be accommodated at the level of `collinearly-improved' Langevin
equations demonstrates that a generalized stochastic picture can still apply beyond leading order,
while at the same time providing a suitable framework for numerical calculations.

\section*{Acknowledgments}
\vspace*{-0.3cm}
We would like to thank Tuomas Lappi for comments on the manuscript and to
Al Mueller for inspiring discussions. E.I is grateful to the Yukawa Institute for Theoretical
Physics (YITP) in Kyoto University for hospitality during the early stages of this work.
This work is supported in part by the France-Japan exchange program Sakura 34205VD.
The work of E.I. is supported by the European Research Council 
under the Advanced Investigator Grant ERC-AD-267258.

\bigskip

\providecommand{\href}[2]{#2}\begingroup\raggedright\endgroup

\end{document}